# Epitaxial graphene quantum dots for high-performance THz bolometers.

*A. El Fatimy[1*], R.L. Myers-Ward[2], A.K. Boyd[2], K.M. Daniels[2], D.K. Gaskill[2], and P. Barbara[1*]*.

[1] Department of Physics, Georgetown University, Washington, DC 20057, USA. [2] U.S. Naval Research Laboratory, Washington, DC 20375, USA.

*e-mail: ae497@georgetown.edu, paola.barbara@georgetown.edu

**Light absorption in graphene causes a large change in electron temperature, due to low electronic heat capacity and weak electron phonon coupling,[1-3] making it very attractive as a hot-electron bolometer material. Unfortunately, the weak variation of electrical resistance with temperature has substantially limited the responsivity of graphene bolometers. Here we show that quantum dots of epitaxial graphene on SiC exhibit an extraordinarily high variation of resistance with temperature due to quantum confinement, higher than 430 MΩ K$^{-1}$ at 2.5 K, leading to responsivities for absorbed THz power above $1 \times 10^{10}$ V W$^{-1}$. This is five orders of magnitude higher than other types of graphene hot electron bolometers. The high responsivity combined with an extremely low noise-equivalent power, about $2 \times 10^{-16}$ W/√Hz at 2.5K, place the performance of graphene quantum dot bolometers well above commercial cooled bolometers. Additionally, these quantum dot bolometers have the potential for superior performance for operation above 77K.**

The electrical resistivity of pristine graphene shows a weak temperature dependence, varying by less than 30% (200% for suspended graphene) from 30 mK to room temperature[4,5], because of the very weak electron-phonon scattering[6]. A stronger temperature dependence was obtained either by using dual-gated bilayer graphene[1,7] to create a tunable band gap[7], or by introducing defects to induce strong localization[2]. Both schemes have successfully produced bolometric detection, with responsivities up to $2 \times 10^5$ V W$^{-1}$ and temperature coefficient for the resistance as high as 22 kΩ K$^{-1}$ at 1.5K [1,2]. These devices required the use of multilayer structures adding complexity. In the case of bilayer graphene, top and bottom gates were needed to electrically induce a bandgap. In the case of disordered graphene, a boron nitride layer was used as a tunneling barrier between the graphene and the electrodes to reduce thermal conductance due to diffusion of the electrons to the electrodes.

Here we demonstrate hot-electron bolometric detection using nano-patterned dots of epitaxial graphene. A bandgap is induced via quantum confinement, without the need of gates, using a simple single-layer structure. We study the THz response of dots with diameter varying from 30 nm to 700 nm, at 0.15 THz and at temperatures from 2.5K to 80K. These devices are extremely sensitive and the responsivity increases by decreasing the dot diameter, with the smaller dots still showing a clear response at liquid nitrogen temperature. Our fabrication process is fully scalable and easily provides multiple devices on the same chip, making it suitable for bolometer arrays. Moreover, its flexibility allows patterning of arrays of dots electrically connected in parallel, to control the device impedance while preserving the strong temperature dependence.

We fabricated our dots using e-beam lithography and a process developed by Yang et al.,[8] (see Methods). Fig. 1a shows an image of a typical quantum dot and the temperature dependence of the resistance for a couple of dots. The resistance varies strongly, by almost four orders of magnitude for the 30 nm dot and two orders of magnitudes for the 150 nm dot over a temperature range of 4 to 300 K; for the case of the 30 nm dots, the temperature coefficient is about 430 MΩ K$^{-1}$ at 6K.

The current-voltage characteristics are non-linear, as shown in Fig. 1b. If we assume that the non-linearity is solely due to Joule heating and the strong temperature dependence of the resistance, we can directly estimate the expected bolometric performance. By using the measured temperature dependence of the resistance and the differential resistance $R = dV_{DC}/dI_{DC}$ as a function of $P_{IN} = V_{DC} \times I_{DC}$ (see Fig. 1c), we can extract the electron temperature *vs.* the electrical power absorbed by the device. The result for a 30-nm dot at the base temperature of 6K is shown in Fig. 1d, yielding a thermal conductance $dP_{IN}/dT_e = G_{TH} \sim 7 \times 10^{-12}$ W K$^{-1}$ in the range 1.5 pW < $P_{IN}$ < 10 pW. We can also estimate the device speed from the thermal time constant, $\tau = C_e/G_{TH} < 2.5$ ns, where $C_e$ is the electronic heat capacity (see SI).

The responsivity *r* is the change in voltage across the device divided by the absorbed power at a fixed bias current. It is directly related to the temperature dependence of the resistance and the thermal conductance, $G_{TH}$, according to

$$r = \frac{\Delta V_{DC}}{\Delta P_{IN}} = I_{DC}\frac{\Delta R}{\Delta P_{IN}} = I_{DC}\frac{\Delta R}{\Delta T}\frac{1}{G_{TH}},$$

We can estimate the expected bolometric responsivity from the plot of resistance as a function of electrical power. For example, by using the $R(P_{IN}) = dV_{DC}/dI_{DC}(P_{IN})$ curve in Fig. 1c, at $P_{IN} = 0.45$ pW, corresponding to $I_{DC} \sim 25$ pA from Fig. 1b, we find $r = 0.65 \times 10^{10}$ VW$^{-1}$ for a 30 nm dot at 6K. This is indeed orders of magnitude higher than any value previously reported for graphene detectors[1-3,9].

The bolometric performance estimated above is based on Joule heating. Next, we measure the bolometric performance of the dots with incident 0.15 THz radiation from a backward wave oscillator (BWO). Fig. 2a shows the response of the same 30-nm dot characterized in Fig.1. The change in the current voltage characteristic due to THz radiation is very similar to the change caused by heating (compare Fig. 1b and Fig. 2a). We extract the absorbed THz power by measuring the electrical power at the point in the dark (radiation off) current-voltage characteristic with the same differential resistance as the zero-bias differential resistance of the current-voltage characteristic with radiation on (red curve), as shown in Fig. 2a, corresponding to about 0.4 pW for this 30-nm dot. (Estimates of the average incident power are in the SI.) The voltage change $\Delta V_{DC}$ at 2.5K due to THz radiation is about 20 mV at 100 pA, yielding, indeed, an extremely high optical responsivity, $r = 5 \times 10^{10}$ VW$^{-1}$, consistent with the high responsivity estimates based on Joule heating.

One important characteristic of bolometric sensors is the noise equivalent power (NEP), which is the lowest detectable power in a 1Hz output bandwidth. Because of this extraordinarily high responsivity, the contribution of Johnson-Nyquist (JN) noise to the noise equivalent power is extremely small, notwithstanding the high device resistance. Other sources of noise that are intrinsic to the bolometer and do not depend on the measurement circuit are shot noise (SN) and thermal fluctuations (TF)[10,11]. We estimate the shot noise from the device $I_{DC}$ at the bias point where we measure the response $\Delta V_{DC}$ and the noise due to thermal fluctuations using the thermal conductance extracted from the electrical characteristics of each device (see discussion above to extract $G_{TH}$ of a 30-nm dot using the measurements in Fig. 1). The total NEP is given by $NEP^2 = NEP^2_{JN} + NEP^2_{SN} + NEP^2_{TF} = (4k_BTR)/r^2 + (2eI_{DC})/r^2 + 4k_BT^2G_{TH}$. The plot of the total NEP as a function of temperature calculated for a 30-nm and a 150-nm dot is in Fig. 2d and the responsivity in Fig. 2c. The devices show remarkable performance up to 77 K, with NEP at least one order of magnitude lower (for the 30 nm dot at 2.5 K) than the best commercial cooled bolometers and much faster response time (a few nanoseconds, compared to milliseconds for commercial bolometers). The 2.5K bolometric performance measured as a function of the dot diameter for several devices is shown in the SI section.

As mentioned above, the strong temperature dependence of the resistance is the key property that produced a dramatic increase of the responsivity. Our quantum dots are made of epitaxial graphene on SiC. The SiC substrate surface has basal plane terraces bounded by steps defined by (11-2n) family of facets. We found that the orientation of the dots with respect to the steps significantly affects their resistance: the devices show higher resistance when the current flows perpendicular to the steps. Anisotropic conductivity due to local scattering at the step edges of the SiC substrates has been observed before and it was proposed that the anisotropy was due to Si atoms trapped at the steps[12].

Fig. 3a shows the temperature dependence of the resistance for dots of different diameter that are patterned with current flowing *perpendicular* to the steps. Regardless of dot orientation and size, the curves could not be fit to a simple function in the temperature range from room temperature to 2.5K.

However, the curves for *all* the dot sizes and orientations show a good fit to thermal activation, in the range 9K < $T$ < 60K, as shown in Fig. 3a. The activation energies extracted from these fits are in Fig. 3b. The black (red) circles correspond to dots with steps perpendicular (parallel) to the direction of the current flow. The activation energies are roughly proportional to the inverse of the dot diameter. This is consistent with quantum dots from exfoliated graphene[13,14], where the combination of the charging energy and the confinement energy open a bandgap that is inversely proportional to the dot diameter. Another notable feature in Fig. 3b is that the black circles clearly show an upward energy shift of about 1 meV. The effect of the substrate on the energy gap of nanostructured graphene has been previously observed in etched graphene nanoribbons, where the disorder potential substantially increased the conductance gap induced by quantum confinement[15]. Here, the steps between the terraces also change the potential landscape of the dots, introducing additional conductance barriers having the same orientation as the steps.

Our physical picture is that, when a bias voltage is applied across the dot, the current is dominated by thermal activation over a potential barrier. When radiation is incident on the bolometer, the electron temperature in the whole graphene area (including the graphene on either side of the dot) increases and the current also increases. Both the quantum confinement gap and the potential barriers from the steps in the substrate contribute to the overall barrier height. The activation energy can therefore be tailored by varying the size of the dot or by applying a gate voltage. Here we do not use a gate electrode; therefore the alignment of the Fermi energy within the confinement gap is not controlled. Nevertheless, the activation energies show a regular dependence on the dot diameter, suggesting that the whole chip is uniformly doped.

Our epitaxial graphene quantum dots show exceptional responsivity and very low NEP, in spite of their relatively simple structure. We are working on devices coupled to antennas[16,17] and new designs tailored to maximize THz absorption in the graphene adjacent to the dot. Future work will also focus on studying Coulomb blockade patterns on gated dots, to separately extract all the characteristic energies and study the effect of the substrate and its orientation with respect to the dot structure.

**Acknowledgments**

This work at Georgetown University was supported by the U.S. Office of Naval Research (award number N000141310865); work at NRL was also supported by the U.S. Office of Naval Research.

**Author Contributions**

Device fabrication and transport measurements were performed by A.E.F.

A.K.B., K.M.D., R.L.M.-W. and D.K.G. synthesized and characterized the graphene on SiC.

A.E.F. and P.B. designed the experiment and analyzed the data. All authors contributed to the discussion of the results and the manuscript preparation.

**Additional Information**

The authors declare no competing financial interests.

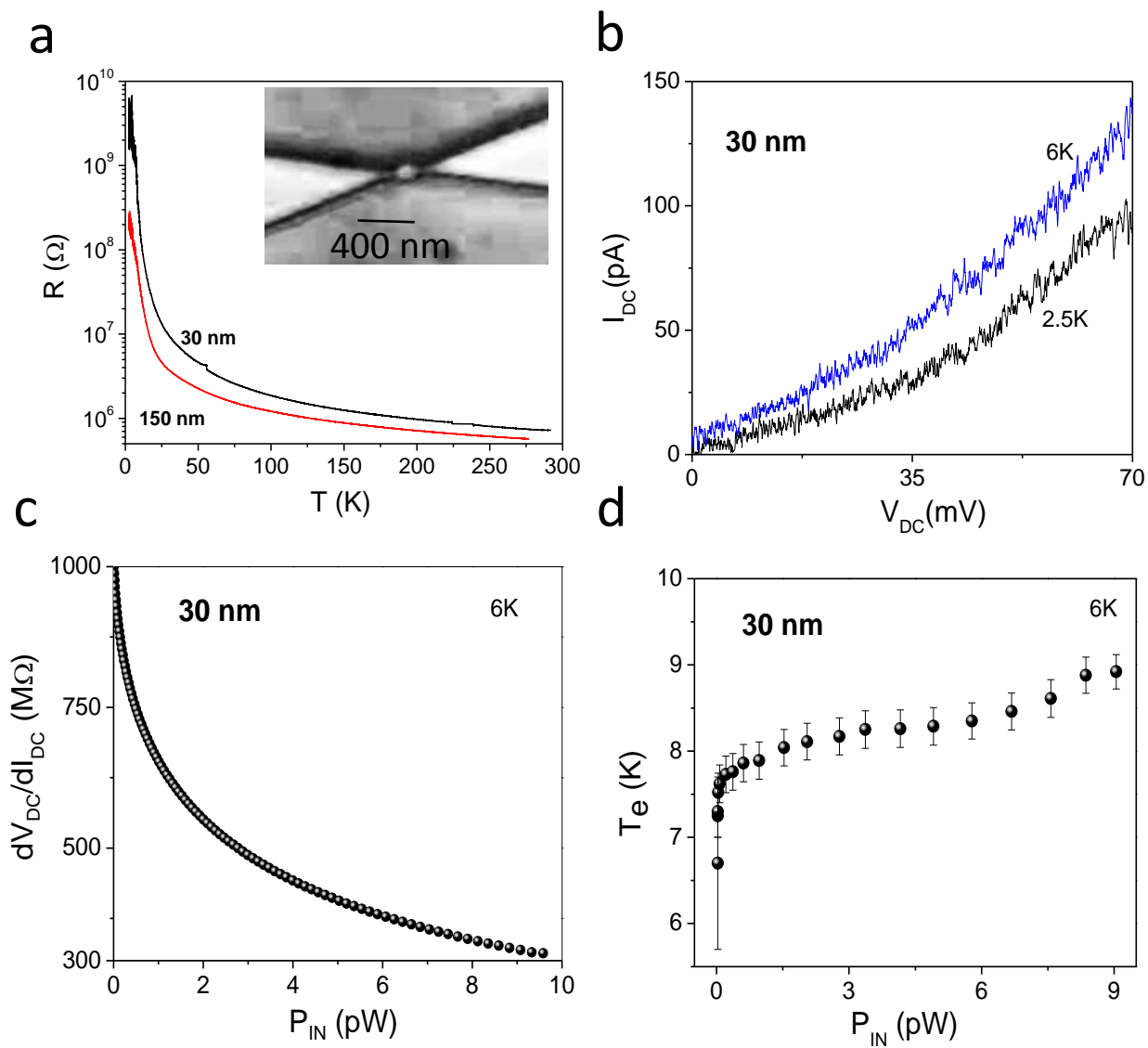

**Figure 1. Temperature dependence and electrical power characterization of graphene quantum dots. a,** Resistance *vs.* temperature for two quantum dots with different diameter at $V_{DC}$ = 5mV. (For $V_{DC}$ < 10 mV, there are no measurable Joule heating and non-linearity of the IV curves.) Inset: Scanning Electron Microscope image of a typical quantum dot. **b,** Current-voltage characteristic of a 30-nm dot at 2.5K and 6K. **c,** differential resistance *vs.* Joule power at 6 K. **d,** Electron temperature as a function of Joule power.

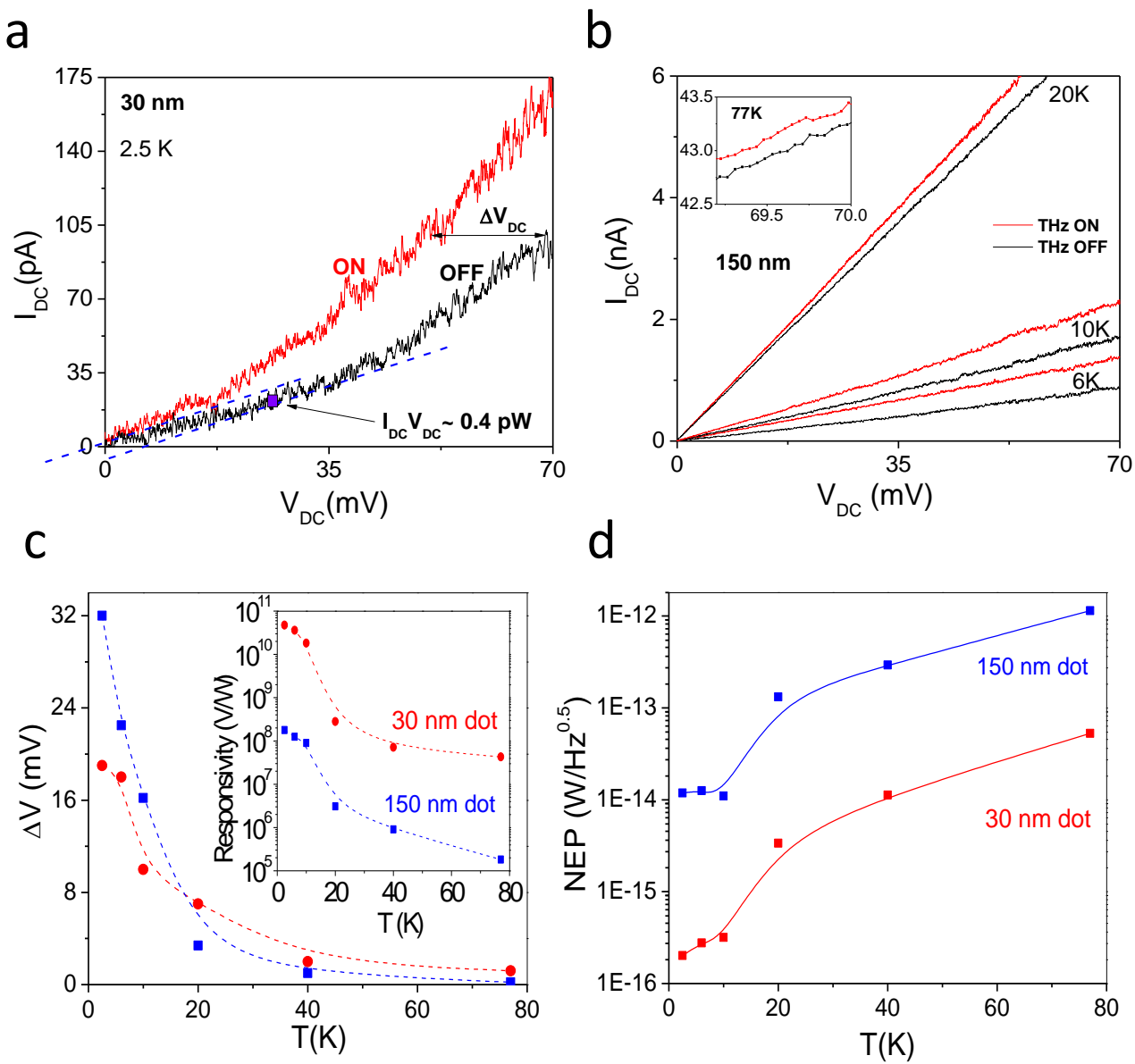

**Figure 2. THz response of graphene dots. a**, Current-voltage characteristic for a 30-nm dot with (red) and without (black) 0.15 THz radiation at 2.5K. The power absorbed from the THz radiation is estimated by finding the Joule power at the point in the black curve with the same differential resistance as the zero-bias differential resistance of the red curve (see the blue dashed lines in **a**). **b**, Response of a 150-nm dot to 0.15 THz radiation at different temperatures. A clear response can still be measured at 70K. **c,** Voltage change and responsivity to 0.15 THz radiation of a 30-nm (red) and a 150-nm (blue) dot at different temperatures. **d,** Calculated noise equivalent power for 30-nm and 150-nm quantum dot bolometers.

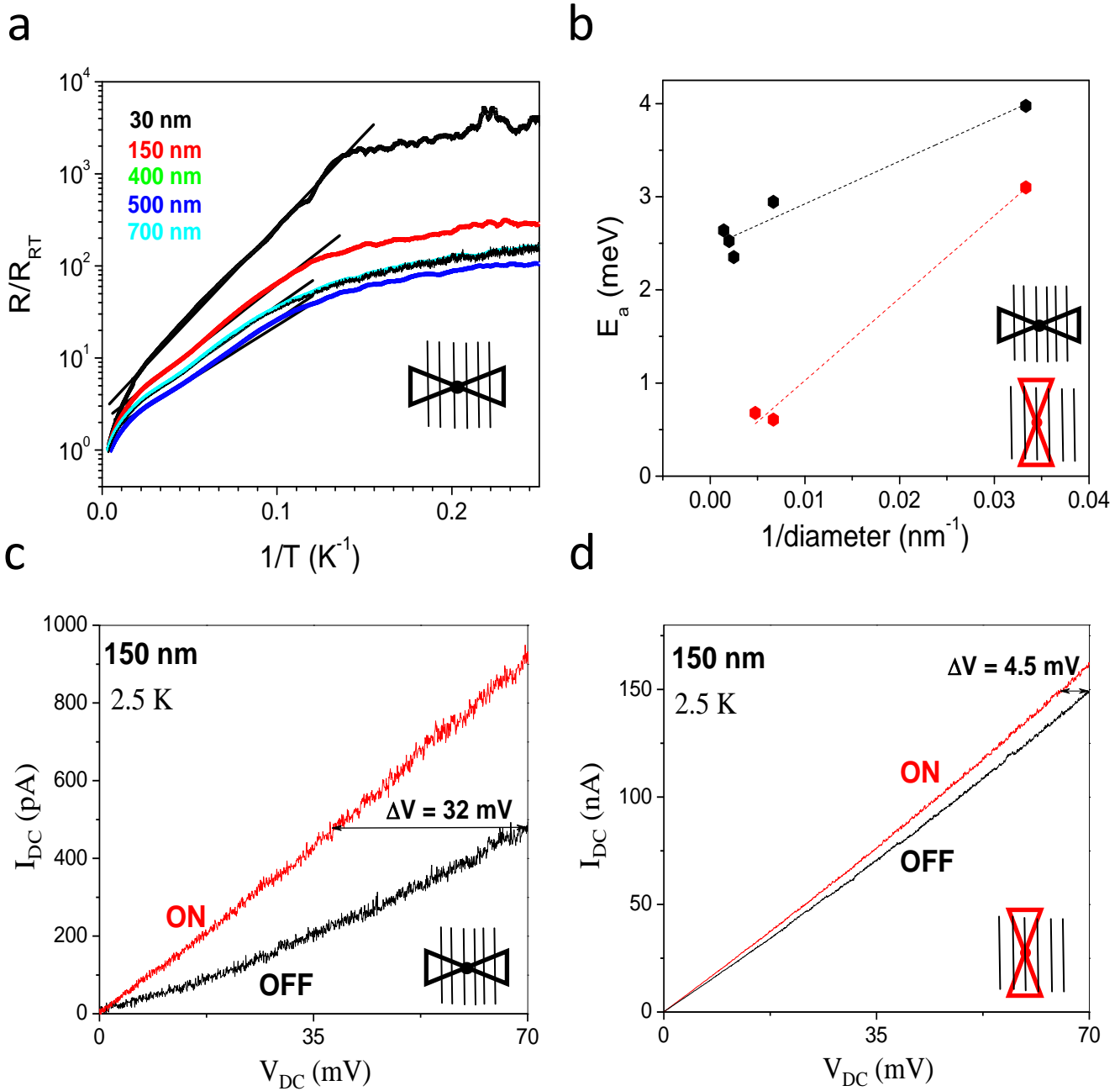

**Figure 3. Temperature dependence and effect of dot orientation with respect to the SiC steps.**
**a,** Resistance normalized to the room temperature value $R_{RT}$ for dots of different size as a function of inverse temperature. The current flow is *perpendicular* to the SiC steps. For each curve we use a linear fit (black lines) in the temperature range showing thermal activation behavior, 9 K < $T$ < 60 K. **b**, Activation energies extracted from the fits in **a** (black circles) and for another set of quantum dots with current flowing *parallel* to the SiC steps (red circles). **c & d,** Current-voltage characteristic for 150-nm dots with (red) and without (black) 0.15 THz radiation at 2.5K. The current flow is perpendicular (parallel) to the SiC steps in **c** (**d**).

**Methods**

Graphene synthesis

Graphene synthesis was accomplished via Si sublimation from semi-insulating (0001) 6H-SiC substrates misoriented *ca.* 0.4° from the basal plane under a 100 mbar Ar pressure in a commercial chemical vapor deposition reactor[18]. The substrates were etched by $H_2$ prior to graphene synthesis. The terraces of the sample were nominally one layer of graphene bounded by steps that were 2 layers. Before device fabrication, the room temperature carrier sheet density and mobility of the 8 mm x 8mm samples shown here were -7.8 x $10^{12}$ $cm^{-2}$ and 790 $cm^2$ $V^{-1}$ $s^{-1}$, respectively.

Device fabrication

We adapted the process developed by Yang *et al.*[8] to use e-beam lithography, instead of photolithography. As a first step, we sputter deposited a 30-nm metallic layer (Pd or Au) directly on graphene. This layer prevents any contamination due to the photo-resist during processing. We spin a polymethyl methacrylate/MMA bilayer on the metal and write the dot pattern by e-beam lithography, with the shape shown in the inset in Fig. 1 (we define the dot pattern as the dot in the center with the two triangular shapes attached to it on both sides). Then we sputter deposit a 50-nm-thick layer of Pd and remove this Pd layer around the dot pattern by lift off, so that the metallic layer on the dot pattern is thicker (30nm + 50 nm) than the metallic layer around it, which is only 30-nm thick. The next step is dry etching of the 30-nm metallic layer around and on top of the dot pattern, using Ar plasma (50 SCCM, 80 W) for 90 seconds. This step leaves about 50-nm thick metal on top of the dot pattern and no metal around it. The metal on the dot pattern works as a mask to protect the graphene underneath in the subsequent steps. Then we etch any graphene left around the dot pattern by $O_2$ plasma (50 SCCM, 10 W) for 60 seconds. Next we pattern the source and drain electrodes by e-beam lithography, followed by sputtering of Cr(3 nm)/Au(150 nm) and lift-off. The 50-nm metallic layer (Pd or Au) covering the dot pattern is removed in the last fabrication step, using aqua regia ($HNO_3$:HCl:$H_2O$ = 1:3:4) for 20 seconds.